\documentclass[twocolumn,iop]{emulateapj}

\newcommand{\noprint}[1]{}

\newcommand{\myemail}{ayw@mso.anu.edu.au}

\newcommand{\kms}{\,\mathrm{km}\,\mathrm{s}^{-1}}

\newcommand{\yr}{\,\mathrm{yr}}

\newcommand{\Kv}{\,\mathrm{K}}

\newcommand{\kpc}{\,\mathrm{kpc}}
\newcommand{\parsec}{\,\mathrm{pc}}

\newcommand{\erg}{\,\mathrm{erg}}
\newcommand{\ergs}{\,\mathrm{erg}\,\mathrm{s}^{-1}}
\newcommand{\cmq}{\,\mathrm{cm}^{-3}}

\newcommand{\msolar}{\,\mathrm{M}_{\sun}}

\newcommand{\vrw}{\left<v_{r,w}\right>}
\newcommand{\vrad}{v_r}
\newcommand{\Pjet}{P_\mathrm{jet}}
\newcommand{\Pjetff}{P_\mathrm{jet,45}}
\newcommand{\Ajet}{A_\mathrm{jet}}
\newcommand{\pjet}{p_\mathrm{jet}}
\newcommand{\vjet}{v_\mathrm{jet}}
\newcommand{\rhojet}{\rho_\mathrm{jet}}
\newcommand{\Gjet}{\Gamma_\mathrm{jet}}
\newcommand{\betajet}{\beta_\mathrm{jet}}
\newcommand{\Mwtot}{M_{w,\mathrm{tot}}}

\newcommand{\nhot}{n_h}
\newcommand{\Thot}{T_h}
\newcommand{\nwarmav}{\left<n_w\right>}
\newcommand{\ncloud}{n_c}
\newcommand{\Rcloud}{R_c}
\newcommand{\rhocloud}{\rho_c}
\newcommand{\tcollapse}{t_c}
\newcommand{\tabl}{t_\mathrm{abl}}
\newcommand{\vabl}{v_\mathrm{abl}}
\newcommand{\pism}{p_\mathrm{ISM}}
\newcommand{\Tcrit}{T_\mathrm{crit}}
\newcommand{\rhocrit}{\rho_\mathrm{crit}}
\newcommand{\Ekinw}{E_{\mathrm{kin},w}}
\newcommand{\phiw}{\phi_w}
\newcommand{\fvol}{f_V}

\newcommand{\Thomsonx}{\sigma_T}
\newcommand{\mprot}{m_p}

\newcommand{\MBH}{M_\mathrm{BH}}
\newcommand{\MBE}{M_\mathrm{BE}}

\newcommand{\etacrit}{\eta_\mathrm{crit}}

\newcommand{\LCDM}{$\Lambda$CDM}

\slugcomment{Submitted to the Astrophysical Journal on \today.}

\shorttitle{Relativistic Jet Feedback}
\shortauthors{Wagner \& Bicknell}

\begin{document}


\title{Relativistic Jet Feedback in Evolving Galaxies}


\author{A. Y. Wagner and G. V. Bicknell}
\affil{Research School of Astronomy and Astrophysics, The Australian National University, ACT 2611, Australia}
\email{\myemail}


\begin{abstract}
Over cosmic time, galaxies grow through the hierarchical merging of smaller galaxies. However, the bright region of the galaxy luminosity function is incompatible with the simplest version of hierarchical merging, and it is believed that feedback from the central black hole in the host galaxies reduces the number of bright galaxies and regulates the co-evolution of black hole and host galaxy. Numerous simulations of galaxy evolution have attempted to include the physical effects of such feedback with a resolution usually exceeding a kiloparsec. However, interactions between jets and the interstellar medium involve processes occurring on less than kiloparsec scales. In order to further the understanding of processes occurring on such scales, we present a suite of simulations of relativistic jets interacting with a fractal two-phase interstellar medium with a resolution of two parsecs and a largest scale of one kiloparsec. The transfer of energy and momentum to the interstellar medium is considerable, and we find that jets with powers in the range of $10^{43}$--$10^{46} \ergs$ can inhibit star formation through the dispersal of dense gas in the galaxy core. We determine the effectiveness of this process as a function of the ratio of the jet power to the Eddington luminosity of the black hole, the pressure of the interstellar medium and the porosity of the dense gas.
\end{abstract}









\section{Introduction}
\label{s:intro}

It is widely believed that feedback from active galactic nuclei (AGN) during the epoch of galaxy formation is required to explain the relation between black hole and bulge mass/velocity dispersion \citep{magorrian98a,gebhardt00a,tremaine02a}, the deficit of bright galaxies in the galaxy luminosity function \citep{cole01a,norberg02a,huang03a}, and the completion of star formation in massive galaxies at epochs of redshift $z \lesssim 2$ \citep{shaver96a,madau96a,bender99a}. It is envisaged that either radiation or outflows from a galactic nucleus impedes the infall of star-forming gas once the central black hole grows to a critical size. Accordingly, \citet{silk98a}, \citet{fabian99a}, \citet{king05a}, and others, have appealed to the physics of either energy-driven or momentum-driven bubbles in order to explain the relationships between the mass of the black hole and the parameters of the host galaxy.

In order to model the galaxy luminosity function, \citet{croton06a} have utilized semi-analytic models based on the output of the Millennium Simulation, incorporating \lq\lq{}radio-mode\rq\rq{} feedback, coupled with a prescription for the accretion rate into the center of each evolving galaxy. Their feedback prescription is motivated by the well-documented evidence for the effect of radio galaxies on \lq\lq{}cooling flow\rq\rq{} galaxies \citep{fabian03a,mcnamara05a}.

There is also a growing literature on cosmological simulations in a \LCDM{} cosmogony involving both dark matter and gas dynamics, which incorporate feedback from both supernovae (SN) and black holes, and which test scenarios of galaxy merging and growth \citep{springel03a,springel03b,booth09a,booth09b,schaye10a}. In simulations using the Gadget-2 code \citep{springel05a} the total number of SPH particles representing the baryonic component exceeds 250 million, and the effective spatial dynamic range is an equally impressive $10^5$ per dimension. Nevertheless, the best spatial resolution is about 2 kpc and does not resolve the spatial scales where important dynamical processes occur. This is highlighted by the prescriptions for black hole growth and feedback described, for example in the work by \citet{booth09b}. Accretion is described in terms of the Bondi-Hoyle accretion rate multiplied by a factor, which can be as large as $10^2$. The rationale for this approach is that at sub-grid scales the density would be larger, and the real accretion rate would be appreciably higher. However, the higher densities and the consequent cooling and fragmentation on sub-grid kpc scales would create a multi-phase interstellar medium (ISM), the physics of which is not satisfactorily captured by the simulations. In particular, such a medium is porous, and the interaction between jets and the dense, potentially star-forming clouds of gas is complex, with radio-emitting plasma being able to channel through holes in the density distribution, rather than isotropically impacting a smooth distribution of dense gas as shown in previous work \citep{sutherland07a}.

Given that black-hole driven feedback occurs in bright galaxies, there does not appear to be a consensus on the type of feedback: How much AGN power is involved, and does the feedback involve radiative or mechanical processes or both? In their radio-mode model, \citet{croton06a} attribute feedback to radio galaxies accreting at rates well below Eddington, and for typical ellipticals, this means low-powered Fanaroff-Riley Class I radio galaxies are the primary drivers of feedback. In \citet{fabian99a} the momentum for dispersing the circumnuclear gas comes from a quasar wind;  in \citet{king05a} the momentum of the outflow is provided by an Eddington-limited radiation-driven wind; in SPH simulations the accretion rate, which is the ultimate power source for an outflow, can approach Eddington values \citep{booth09a}.

There is also an issue of what class of AGN actually drives black hole feedback. The models by \citet{croton06a} invoke low-powered radio galaxies. Observers, however, have focused on powerful radio galaxies, mainly at $z\gtrsim 2$; in these galaxies there is evidence for substantial outflows of line-emitting gas and neutral gas driven by the radio jets \citep{nesvadba09b, morganti10a}. It is possible that there is a role for radio galaxies with a range of powers: Powerful radio galaxies may be responsible for the establishment of the stellar mass at $z\gtrsim 2$, and less powerful sources may be responsible for the maintenance of the stellar content through the inhibition of cooling flows \citep{nulsen09a}. We also note that radio galaxies are mainly relevant to the elliptical galaxy population, and that the separate luminosity functions for early and late-type galaxies \citep{huang03a} indicate the requirement for feedback in both populations.

In this paper we consider the potential role of radio galaxies in AGN feedback and address the following questions: (1) What jet power is required for the radio galaxy phase to have an important effect on inhibiting star formation in a given host, and (2) Is the range of radio powers broad enough that radio galaxies could affect the entire distribution of bright ellipticals? In this paper we present progress in answering these questions through simulations with a resolution of $2\parsec$ per computational cell. These simulations confront the sub-grid physics that current large scale SPH simulations do not address. In particular, we consider the effect of powerful relativistic jets on a two-phase ISM consisting of hot gas, in which is embedded a dense porous phase of warm gas.

These simulations extend the simulations described in \citep{sutherland07a}, in which a jet with a kinetic power of $3 \times 10^{43} \ergs$ propagates through an inhomogeneous medium in the form of an almost Keplerian fractal disk. It is evident from that simulation that in the geometry considered, jets of that power could not exert enough impact on the clouds to disperse them, and that the jets would not have a important effect on star formation in the core of the host galaxy. In the present simulations we consider jets with powers ranging from $10^{43}$ to $10^{46} \ergs$ propagating through a two-phase medium, in which the dense clouds are spherically distributed throughout a region $1\kpc$ in diameter. These initial data are meant to describe a typical protogalaxy, in which dense gas has accumulated in the core. The fractal distribution of the dense gas enables us to directly examine the effect of porosity on the evolution of potentially star-forming clouds.

In the following sections we describe the parameters of the simulations in more detail and then discuss our results.

\section{Model parameters and initial conditions}

In our simulations, we use the publicly available, open-source code FLASH \citep{Fryxell-etal2000} version 3.2, to which we have added code to incorporate radiative cooling of thermal gas and code to advance advected scalars in the relativistic hydrodynamic solver. Details of the solver are in \citep{MB2005,MPB2005}. We exploit the adaptive mesh capabilities of FLASH, utilizing up to seven levels of refinement; this corresponds to a nominal cubical simulation grid of $1\kpc^{3}$ in physical dimensions, consisting of $512^3$ cells at maximum resolution. However, note that a restricted one parameter scaling of physical dimensions is possible \citep{sutherland07a}.

The relevant jet parameters, which are initiated and maintained constant at the boundary that is the jet inlet, are: the Lorentz factor $\Gjet = (1-\betajet^2)^{-1/2} $, where the  $\betajet = \vjet/c$ is the jet velocity in units of the speed of light; the proper density parameter, $\chi = (\gamma -1)\,\rhojet c^2/ \gamma \pjet$, where $\gamma$ is the polytropic index, $\rhojet$ is the rest mass density, and $\pjet$ is the pressure; and the jet power. Let $\Ajet$ be the jet cross-sectional area. The jet power is 
\begin{equation}
  \Pjet=\frac{\gamma}{\gamma-1}c\pjet \Gjet^2 \betajet \Ajet \left(1+\frac{\Gjet-1}{\Gjet}\chi\right)\,. \\
\end{equation}
We adopt $\gamma=5/3$ for both jet and ambient gas. In all simulations presented here, $\Gjet=10$, and $\chi=1.6$. The jet inlet is a circular region of area $\Ajet$ centered at $(x,y,z)=(0,0,0)$ with normal $(1,0,0)$. The initial jet velocity is parallel to the $x$-axis. The boundary $x=0$ is reflective, apart from the jet inlet. All other boundaries of the simulation domain are designated as inflow/outflow boundaries.

The two parameters describing the hot phase of the ISM are the temperature, $\Thot$, which is fixed at $10^7 \Kv$, and the total number density $\nhot$. The pressure, $p/k$, where $k$ is Boltzmann's constant, for the ISM in each of the simulations is either $10^6$ or $10^7$, typical of the ISM in giant elliptical galaxies \citep{MB2003}. 

The distribution of dense, warm clouds is prescribed in similar fashion to that in \citet{sutherland07a}, following work on terrestrial clouds by \citet{LA2002}. The main difference is that, here, the average mass density is uniform, whereas in \citet{sutherland07a} the average density is that of a near-Keplerian disk. For completeness, we describe the other details of the cloud distribution used in this series of simulations: 
\begin{enumerate}
\item We begin by constructing a cube of density fluctuations in which the single point statistics of the density distribution are described by a log-normal distribution. The mean, 
$\mu$, of the parent distribution is 1.0, and the variance,
$\sigma^2 = 5.0$, these values being consistent with starburst reddening and extinction models \citep{fischera03a,dopita04a}.
This value of $\sigma^2$ is the same as that used in \citet{sutherland07a}. In numerical simulations of supersonic turbulence, \citet{Federrath-etal2010} find $\sigma^2 \approx 3.6$ and 35 for solenoidal (divergence free) and compressive (curl free) forcing respectively, so that our adopted value of 5 is closer to their solenoidal result.

\item The power spectrum of the density distribution $D(k) \propto k^{-5/3}$ for the range of waves numbers $k_{\rm min} < k < k_{\rm max}$ and zero outside of this range.
The parameter $k_{\rm min}$ sets the maximum cloud size. For these simulations $k_{\rm min}=20$ (in cell units in Fourier space) and this value limits the maximum size of an individual cloud to approximately $25\parsec$ for 512 cells along each axis of the $1\kpc$ computational cube. Limiting the cloud size to $25\parsec$ allows for appropriate coverage and variation of hot phase ISM and warm phase ISM along any path through the region of our simulation grid filled with clouds. This ensures that the spatial distribution of clouds is approximately isotropic, and that the jet plasma encounters a statistically significant numbers of clouds along its main axis of propagation. As a result, the flow of the jet plasma through the hot phase is approximately isotropic. The parameter $k_{\rm max}$ is set by the resolution of the simulation. Here $k_{\rm max}=255$ and is equivalent to two computational cells.

\item The parent, unit mean distribution is scaled by the mean particle density of warm clouds, $\nwarmav$; this parameter is determined by the ratio of $\nwarmav$ to $\nhot$.   

\item The temperature in each cell of the warm gas distribution is determined by pressure equilibrium with the hot gas. When the warm gas temperature exceeds $T_{\rm crit} = 3 \times 10^4 \> \rm K$ (in the lowest density parts of the distribution), it is deemed to be thermally unstable and is replaced by hot gas. This makes the gas porous. We can then define a volume filling factor, $\fvol$, of the warm phase by integrating over the log-normal density distribution above the critical density $\rhocrit=\bar{\mu} \, u\pism/\Tcrit{}k$, where $\bar{\mu}\approx0.6156$ is the mean molecular weight and $u$ is an atomic mass unit \citep[see also Appendix~B of][]{sutherland07a}.

For a given value of $\Tcrit$, the volume filling factor is determined by the ratio of the mean warm phase density to hot phase density, $\nwarmav/\nhot$; the larger this parameter the larger the filling factor. This log-normal, fractal, porous distribution of clouds is an important feature of the simulations presented here. 

\citet{Silk1997,Silk2001} and \citet{SN2009} identified porosity as a key parameter in their models of feedback-regulated star-formation, since the porosity determines the extent to which energy-driven bubbles generated by SN or AGN activity are confined. The porosity in the two-phase ISM employed in our simulations plays a similar role, in that it constrains the progress of the jet along its principal axis and ensures the confinement of the pressurized bubble for times much larger than the dynamical time of an unimpeded jet.

The porosity of dense clouds is the primary reason for the differences in the evolution of the radio source and the differences in energy and momentum imparted by the jet to the ISM between our simulations and those of a jet propagating into a uniform medium.

\item The radial extent of the gas is truncated by a sphere of radius $0.5\kpc$. Hence, we are simulating the interaction of jets with warm gas within 1--2 core radii of the parent galaxy. This justifies the neglect of gravity.

\item We include non-equilibrium, optically thin atomic cooling for $T>10^4\Kv$ \citep{sutherland93c,sutherland03b}; for $T \leq 10^4\Kv$ the cooling is set to zero. We use updated solar abundances \citep{AGS2005}.

The non-equilibrium, cooling function is pre-calculated for a high temperature ($10^{10}\Kv$) shock. The main reason for using such a function was outlined in \S~2.1 of \citep{sutherland03b} and, for completeness, we summarize it here: Following a shock there is a short-lived cooling spike as low-ionization material is suddenly shocked to a high temperature. In a numerically well-resolved shock this phase is short-lived and provides about 1\% of the total cooling. However, in a simulation in which each shock is not well-resolved (as is the case here) this spike can dominate the cooling and the situation is exacerbated by the interpolation of intermediate temperatures at each cell. Apart from the initial (and unimportant) cooling spike, our adopted non-equilibrium cooling function is a good approximation to the cooling function of shocks for velocities above $150\kms$. Since the cooling function is pre-calculated and interpolated within the code there are no additional computational costs associated with a non-equilibrium calculation.
\end{enumerate}

The utilization of an \emph{optically thin} cooling function also requires some comment, given the dense, reasonably extended regions that evolve within the simulations. Consider  a region with electron density $n_e= 10^3 \, n_{e,3}\cmq$ and thickness $l$. The electron scattering optical depth $\tau_e \approx 2.0 \times 10^{-3} \, n_{e,3} (l/\rm pc)$. In the simulations with the densest gas the initial average electron density $n_{e,3} \sim 1$ so that the largest clouds with $l \sim 50\parsec$ would be optically thin but verging on optically thick where the density is a factor of 10 above average. As a result of the radiative shocks, the electron density increases by a factor of $10-100$. Hence, even some of the less dense regions could become optically thick to scattering. Scattering alone is not enough to invalidate the assumption of optically thin cooling, and an admixture of dust is required to cause absorption. Moreover, the cooling in such regions occurs before maximum density has been obtained, and photons emitted by cooling plasma following a radiative shock have an escape route through the lower density regions of the shock. In addition, the temperature of the gas in the cooling, optically thin region of a radiative shock would be so high that dust would be destroyed. Hence, optically thin cooling is a reasonable first approximation, which \emph{may} be limited in the largest clouds. More refined radiation hydrodynamic simulations, possibly involving larger clouds, may need to take optical depth effects into account.

\begin{deluxetable*}{ccccccc}
\tablecaption{Simulation parameters\label{t:parameters}}
\tablewidth{0pt}
\tablehead{ 
\colhead{Simulation} & \colhead{$\log \, \Pjet$\tablenotemark{(a)}} & 
\colhead{$\nhot$\tablenotemark{(b)}} & \colhead{$\pism/k$\tablenotemark{(c)}} & 
\colhead{$\nwarmav$\tablenotemark{(d)}} & \colhead{$\fvol$\tablenotemark{(e)}} & \colhead{$\Mwtot$\tablenotemark{(f)}}\\
\colhead{} & \colhead{$(\erg)$} & \colhead{$(\cmq)$} & \colhead{$(\cmq\Kv)$} &
\colhead{$(\cmq)$} & \colhead{}  & \colhead{$(10^{9}\msolar)$}
}
\startdata
A & 45 & 0.1 & $10^6$ & \nodata & \nodata &  \nodata \\
B & 46 & 1.0 & $10^7$  & 1000 & 0.42 & 16 \\
C & 46 & 0.1 &  $10^6$ & 100 & 0.42 & 1.6 \\
C$^\prime$  & 46 & 0.1 &  $10^6$ & 30 & 0.13 & 0.32 \\
D & 45 & 1.0 & $10^7$  & 1000 & 0.42 & 16 \\
D$^\prime$  & 45 & 1.0 & $10^7$  &  300 & 0.13 & 3.2 \\
E & 45 & 0.1 &  $10^6$ & 100 & 0.42 & 1.6 \\
E$^\prime$  & 45 & 0.1 &  $10^6$ & 30 & 0.13 & 0.32 \\
F & 44 & 0.1 & $10^6$ & 100 & 0.42 & 1.6\\
F$^\prime$  & 44 & 0.1 & $10^6$ & 30 & 0.13 & 0.32\\
G & 44 & 1.0 & $10^7$  & 1000 & 0.42 & 16 \\
G$^\prime$  & 44 & 1.0 & $10^7$ & 300 & 0.13 & 3.2\\
H & 43 & 0.1 & $10^6$ & 100 & 0.42 & 1.6
\enddata
\tablecomments{Runs labeled with primed (\lq\lq$\,^\prime\,$\rq\rq) letters denote lower filling factor counterparts to runs with the same letter.}
\tablenotetext{(a)}{Jet power}
\tablenotetext{(b)}{Density of hot phase}
\tablenotetext{(c)}{$p/k$ of both hot and warm phases}
\tablenotetext{(d)}{Average density of warm phase}
\tablenotetext{(e)}{Volume filling factor of warm phase}
\tablenotetext{(f)}{Total mass in warm phase}
\end{deluxetable*}

In summary, the key model parameters used in the different simulations are the jet power and the densities of the hot and warm ISM phases, which determine the cloud filling factor. Table~\ref{t:parameters} summarizes the assigned and derived parameters that we used for our simulations.

\section{Results}
\label{s:results}
\subsection{Morphology of radio source and interstellar medium}

\begin{figure*}[h!]
  \epsscale{1.0}
  \plotone{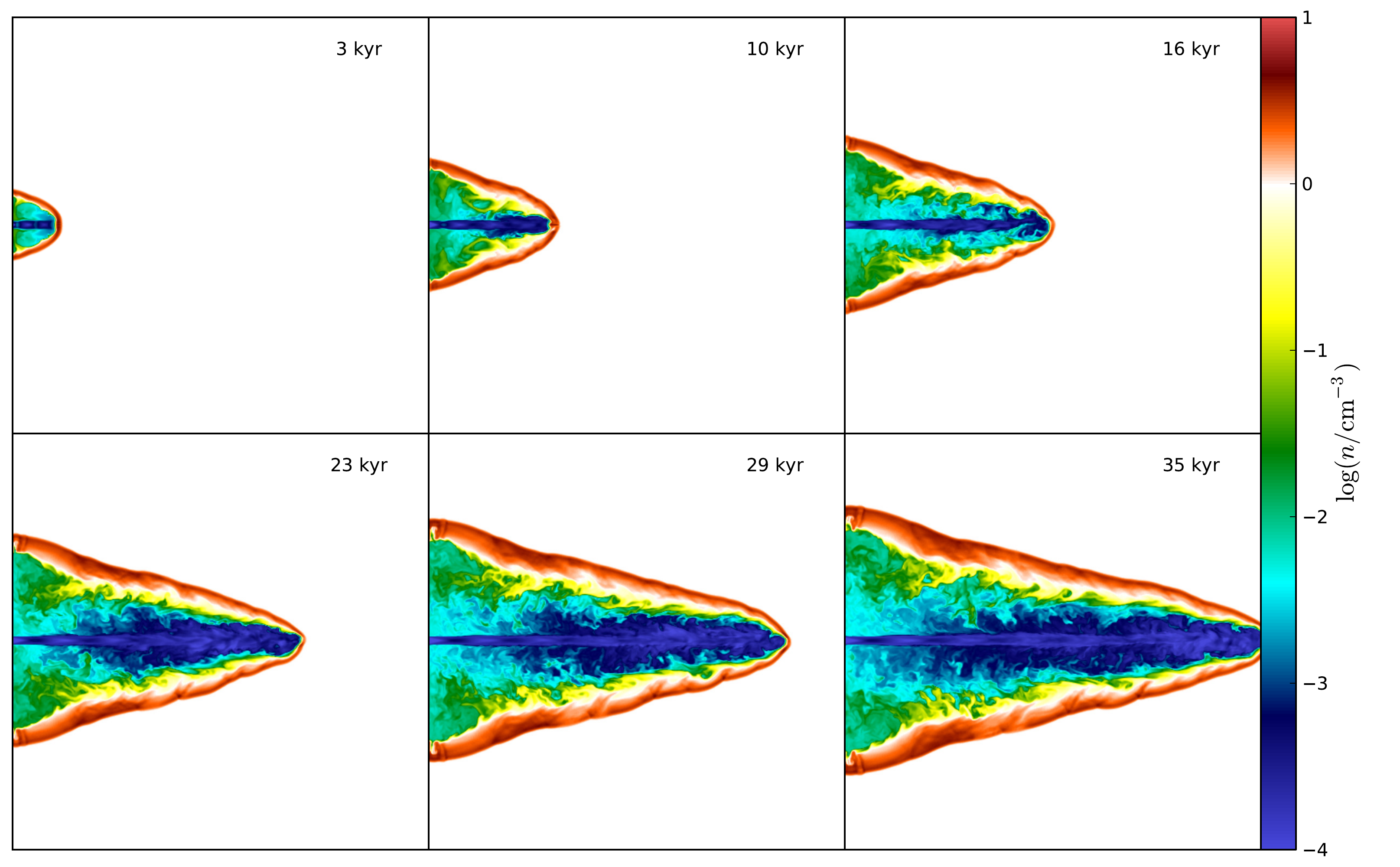}
  \caption{Evolution of density in simulations A. The width and height of each panel are $1\kpc$. This figure is also available as an mpeg animation in the electronic edition of the {\it Astrophysical Journal}.}
  \label{f:densityA}
\end{figure*}
\begin{figure*}[h!]
  \figurenum{2}
  \epsscale{1.0}
  \plotone{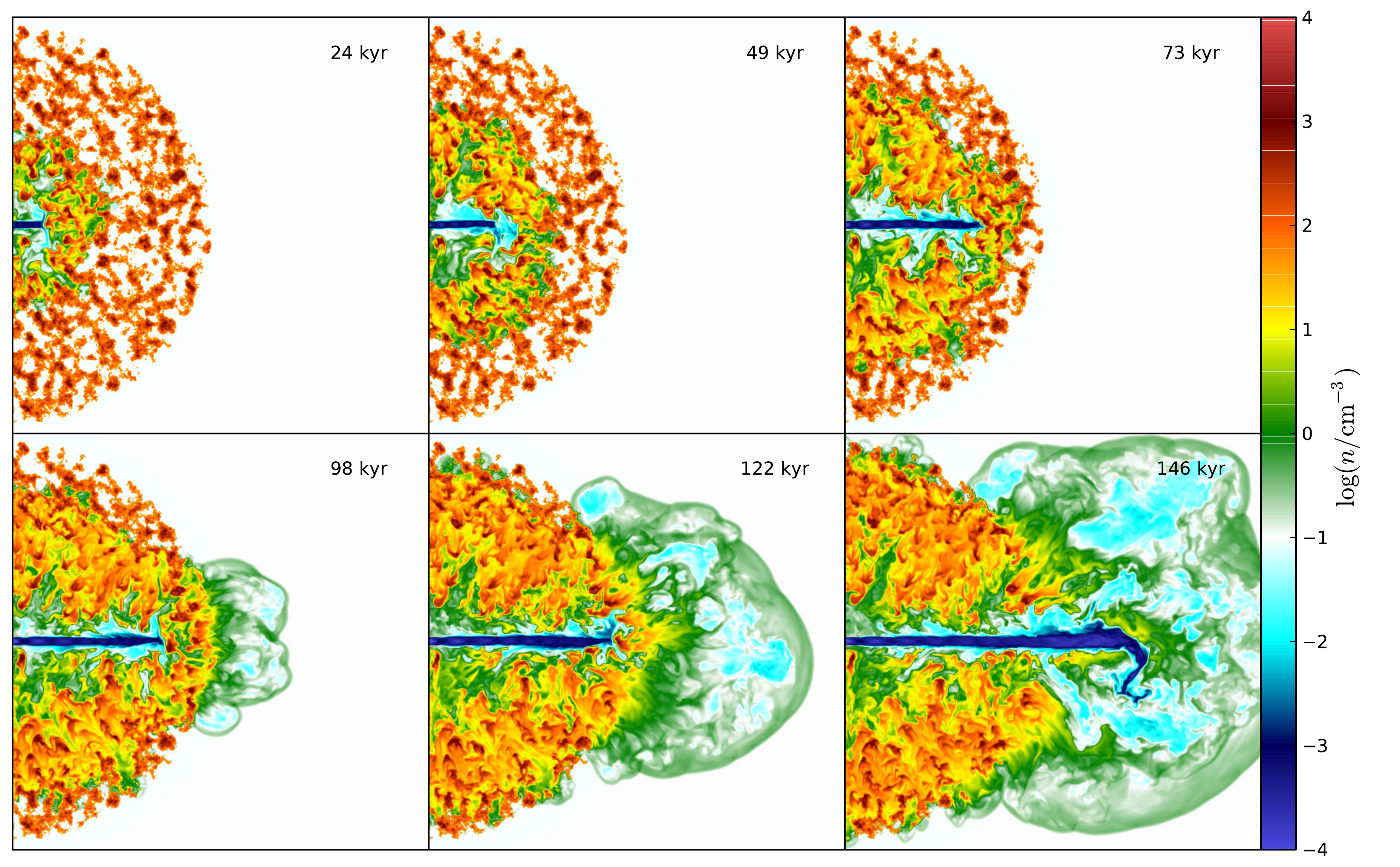}
  \caption{Evolution of density in simulation E. The width and height of each panel are $1\kpc$. This figure is also available as an mpeg animation in the electronic edition of the {\it Astrophysical Journal}. Figures~2.1--2.5 and corresponding mpeg animations depicting the evolution of the density in runs E, C$^\prime$, D$^\prime$, G, and H, respectively, are also available in the electronic edition of the journal.}
  \label{f:densityE}
\end{figure*}

The panels in Figs.~\ref{f:densityA} and \ref{f:densityE} show the evolution of the density in the two simulations A and E, respectively. The figures present slices through the mid-plane $z=0$ of the grid. A is a reference simulation of a jet and homogeneous ISM.  The progress of such a jet has been well established  over two decades of research, although we do note the unstable jittering of the three-dimensional jet once it has traversed about 0.5~kpc (\citealp{MHN2007}, also evident in \citealp{sutherland07a}), which leads to multiple hot spots. 

The progress of the jet in Fig.~\ref{f:densityE} (simulation E) is completely different: Initially the jet is deflected in various directions as it floods through the porous screen of dense clouds, finding channels of least resistance and gradually dispersing the clouds through the effect of the ram pressure of the non-thermal plasma. In addition, the slower rate of progress of the jet traps the high pressure cocoon material; a pseudo-spherical bubble is driven into the ISM, and the dense clouds are driven outwards and dispersed as this bubble expands. Some of the dispersed material is accelerated to speeds $\sim1000\kms$ while the speed of the densest parts of the clouds reaches several hundred $\kms$.

The lateral extent of the cocoon of the A and E jets is very different. When the cloud-free jet has progressed about 0.4 kpc its radial extent is about 0.2 kpc, whereas the radial extent of the cocoon from the cloud-impeded jet is about 0.4~kpc. For the same length the latter jet processed a factor of 8 larger volume of the ISM, removing one of the reservations about the relevance of powerful radio galaxies to AGN feedback. Note also the streams of low density, pressurized cocoon material shooting out in random directions. This is a direct result of the porous ISM, into which the main jet is propagating. Figures~2.1--2.5 and corresponding mpeg animations depicting the density evolution of runs E, C$^\prime$, D$^\prime$, G, and H, respectively, are available in the electronic edition of the {\it Astrophysical Journal}.

\begin{figure*}
  \epsscale{1.00}
  \figurenum{3}
  \plotone{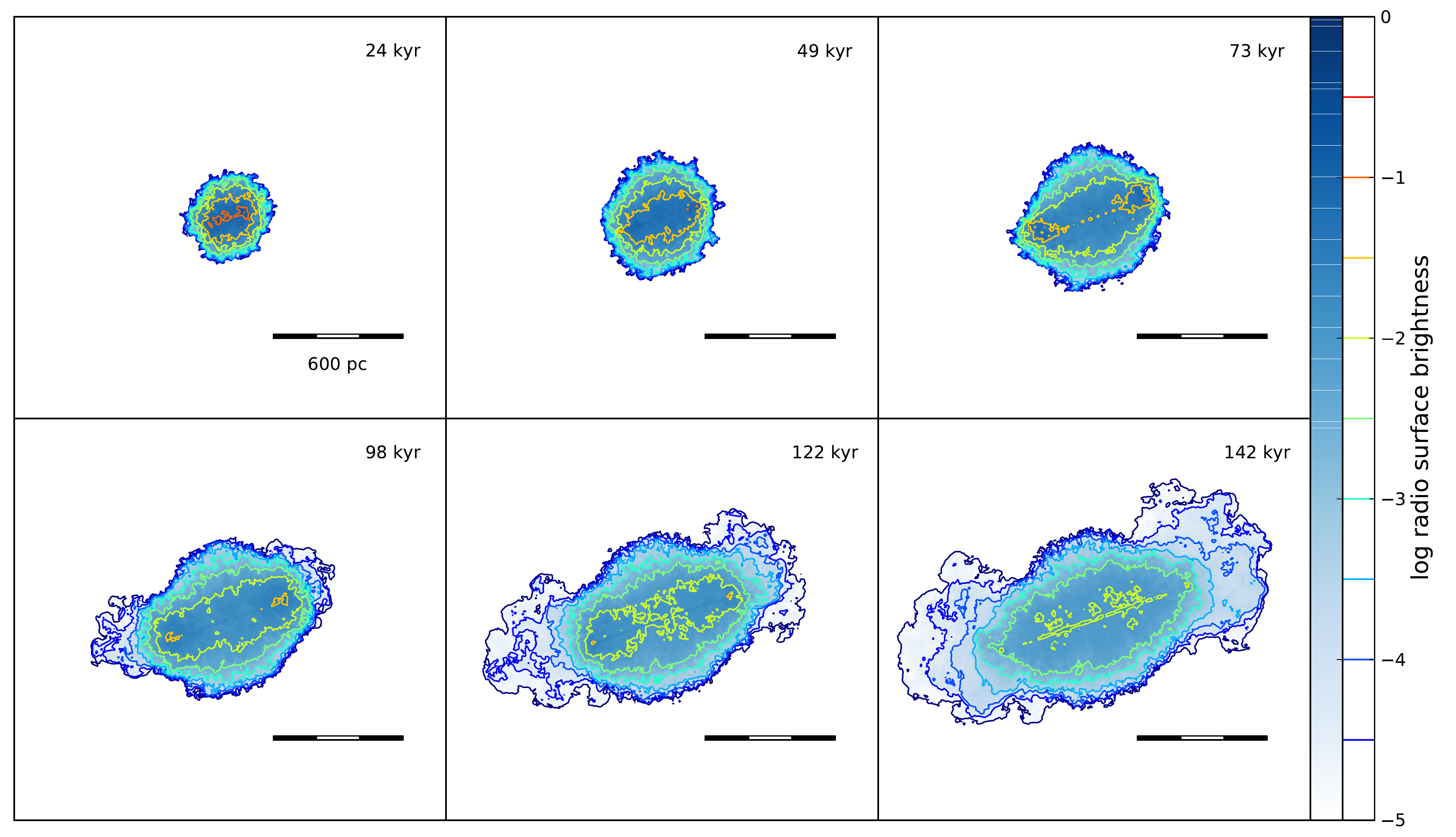}
  \caption{Synthetic radio surface brightness for simulation E at various epochs as indicated.}
  \label{f:surfbrite}
\end{figure*}

We have generated images of synthetic radio surface brightness, based on the density of the non-thermal plasma and also assuming that the magnetic field is proportional to the pressure and a spectral index of 0.6. A set of snapshots from simulation E are shown in Fig.~\ref{f:surfbrite}. The interaction between jet and clouds produce a markedly different morphology from that of a classical radio galaxy. Images such as these can be informatively compared with images of young GPS and CSS radio galaxies \citep[cf.][]{sutherland07a}.

\subsection{Velocity of dispersed clouds}

A convenient approach for assessing the effects of AGN feedback on galaxy formation is to examine the velocity imparted to gas, which could form new stars. The usual criterion for inhibition of further galaxy formation is that the velocities imparted to the clouds is greater than the velocity dispersion, $\sigma$, of the host galaxy \citep{silk98a,king05a}. This does not necessarily mean that the clouds would be ejected to large radii but it does mean that the clouds would be highly dispersed within the potential well of the host galaxy. 

\begin{figure*}
  \figurenum{4}
  \epsscale{1.00}
  \plotone{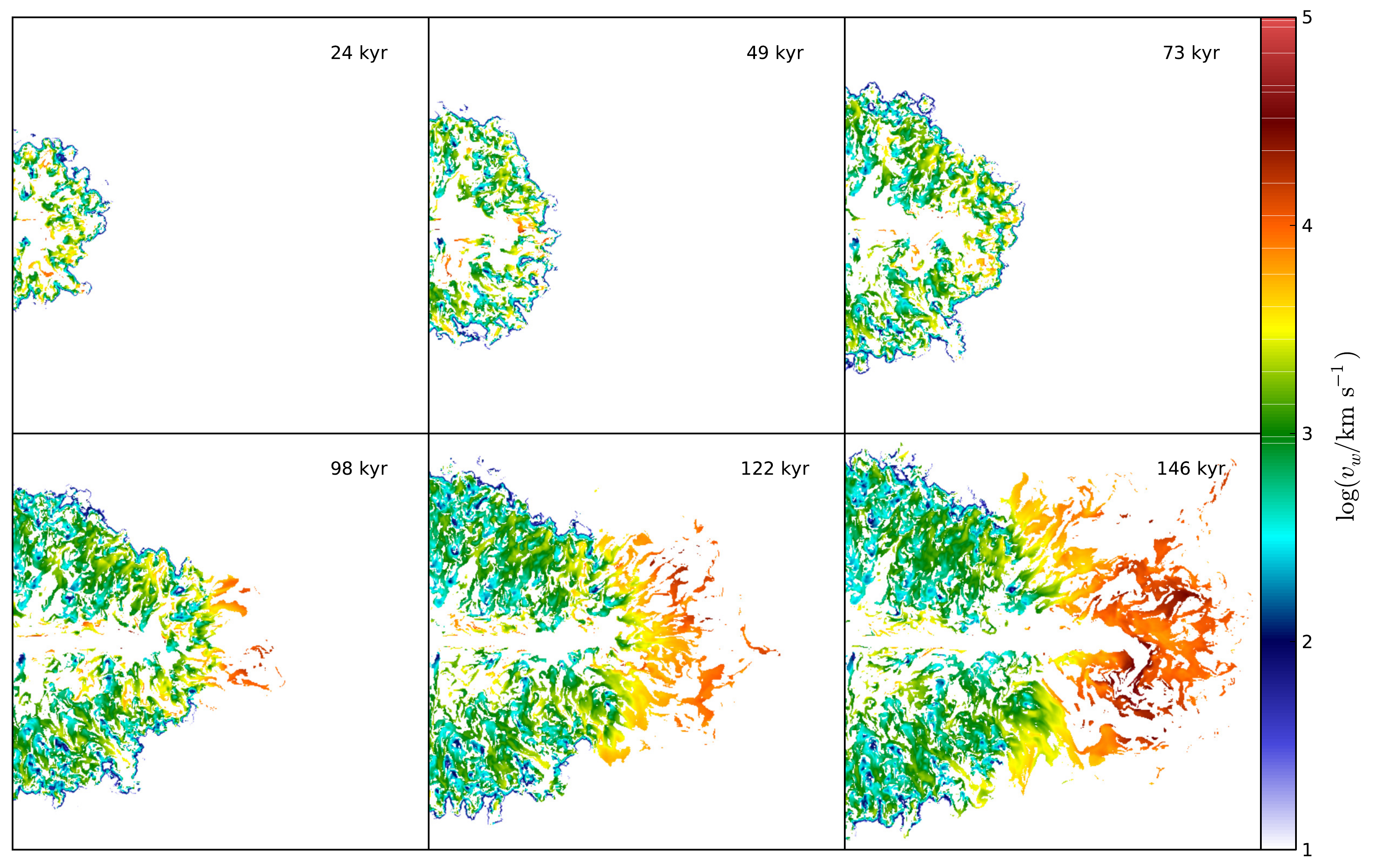}
  \caption{Speed of the warm gas for the value of the warm gas tracer $\phiw>0.9$ in simulation E. The width and height of each panel are $1\kpc$. This figure is also available as an mpeg animation in the electronic edition of the {\it Astrophysical Journal}. Figures~4.1--4.5 and corresponding mpeg animations depicting the evolution of the speed of the warm gas in runs E, C$^\prime$, D$^\prime$, G, and H, respectively, are also available in the electronic edition of the journal.}
  \label{f:v_clouds}
\end{figure*}

In order to track the different gas components we use tracers, which are the mass concentration of that component in each cell. In particular we use a warm gas tracer, $\phiw$, which is initialized to unity in each cell of warm thermal gas.  Figure~\ref{f:v_clouds} shows the velocity of the dispersed warm interstellar material, for $\phiw>0.9$, in simulation~E at various epochs. At $t \approx 98 \, \rm kyr$ the maximum mean cloud velocity is reached. There is a wide range in velocity with the remaining dense cloud cores showing the lowest velocities, material resulting from the dispersal of the clouds showing intermediate velocities $\sim 300\,\rm km\,s^{-1}$, and lighter material showing even higher velocities $\sim 1000 \kms$. Figures~4.1--4.5 and corresponding mpeg animations that depict the velocity evolution of the warm material, for which $\phiw>0.9$, in runs E, C$^\prime$, D$^\prime$, G, and H, respectively, are available in the electronic edition of the {\it Astrophysical Journal}.

Taking $\rho$, the density in a cell, and $v_r$, the radial velocity in a cell, we compute for each simulation the mass-weighted mean radial velocity over all $N$ cells defined by
\begin{equation}
\vrw = \frac {\sum\limits_{l=1}^N \phiw \, \rho \, \vrad}{\sum\limits_{l=1}^N \phiw \, \rho}\,.
\end{equation}
We adopt $\vrw > \sigma$ as the criterion for inhibition of further galaxy formation. This raises the question: What value of $\sigma$ is appropriate? If we use a value relevant to a giant elliptical $\sigma \sim 300 \kms$, this discriminates against lower mass galaxies ($\sigma \sim 200 \kms$ say), in which jet inhibition of star formation may be important in earlier epochs of galaxy formation. In order to establish the relevance of these simulations for all phases of galaxy formation we adopt the following procedure, in which the jet power is parametrized by its ratio $\eta$ relative to the Eddington luminosity, that is, $\Pjet = \eta \, 4 \pi G \MBH \mprot c\Thomsonx^{-1}$, where $\mprot$ is the proton mass and $\Thomsonx$ is the Thomson scattering cross section.
Adopting the Magorrian relation \citep{tremaine02a} between black hole mass and velocity dispersion,
$
{\MBH} \approx 8.1 \times 10^6 \, \sigma_{100}^4 \msolar
$,
where $\sigma_{100}$ is the velocity dispersion in units of $ 100 \kms$, gives the jet power
$
\Pjet = 1.02 \times 10^{45} \, \eta \, \sigma_{100}^4 \ergs
$,
and the velocity dispersion expressed in terms of the jet power in units of $10^{45}\ergs$ is
\begin{equation}
\sigma_{100} \approx 1.0 \, \eta^{-1/4} \, \Pjetff^{1/4}\,.
\label{e:sigma-P}
\end{equation}

\begin{figure*}
  \epsscale{0.8}
  \figurenum{5}
  \plotone{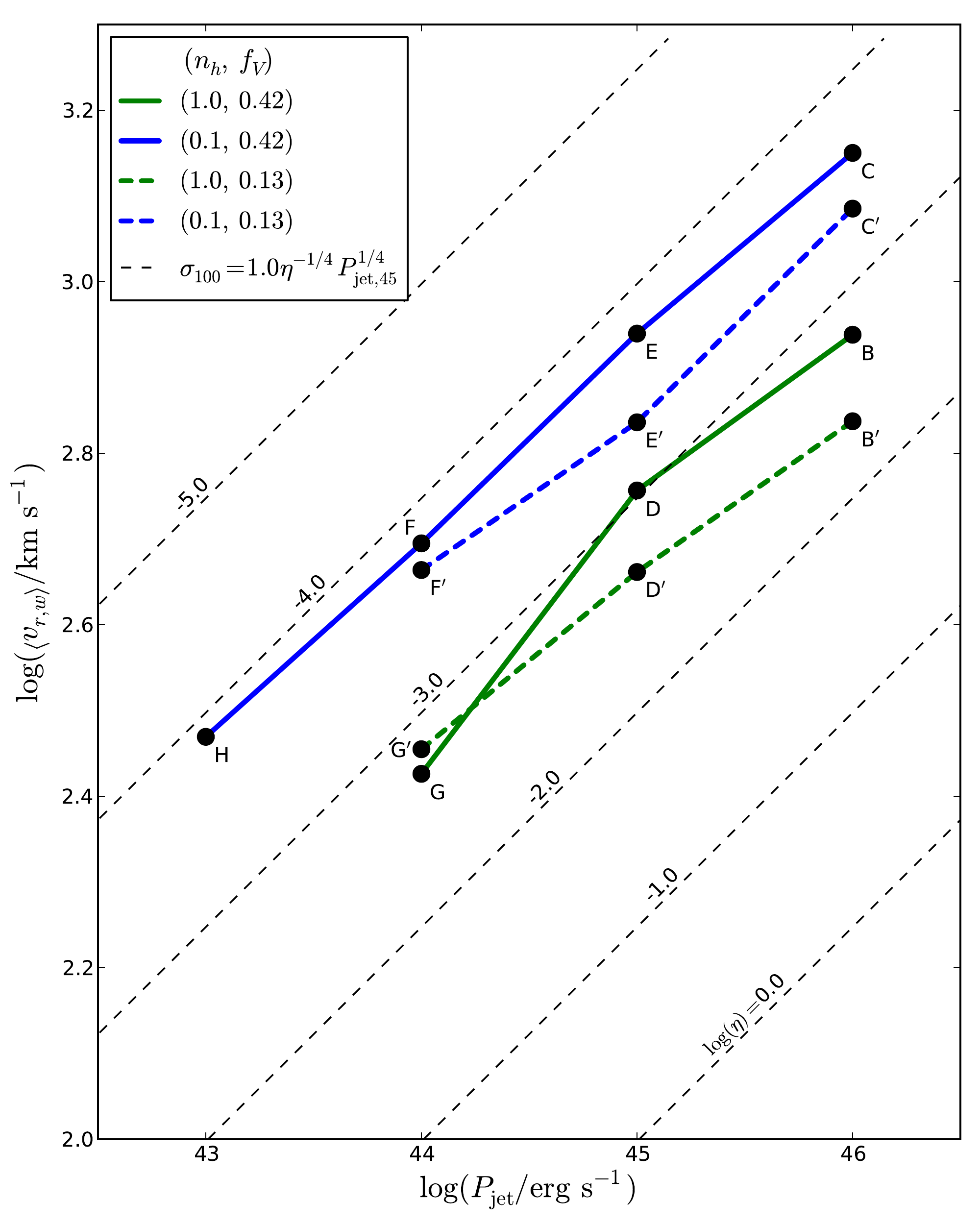}
  \caption{Maximum mean radial velocity of clouds against jet power for the simulations B--H of Table~\ref{t:parameters}. The loci of constant $\eta$, the ratio of jet power to Eddington luminosity are superposed.}
  \label{f:v-P}
\end{figure*}

We plot the results from each simulation on a diagram, Fig.~\ref{f:v-P}, of the maximum value of $\vrw$ vs $\Pjet$. Also plotted are the loci of $\sigma$ vs $\Pjet$ for incremental values of $\log_{10} \eta$, using equation~(\ref{e:sigma-P}).  For a given value of $\eta$ a jet will disperse the warm cloudy material if the corresponding $(\vrw, \Pjet)$ point lies above the specific $\eta = \mathrm{constant}$ locus. For example, consider the point for simulation F represented by the point $\Pjet = 10^{44} \ergs$, $\vrw= 500 \kms$. If $\eta=10^{-2}$, the cloud speed comfortably exceeds the corresponding velocity dispersion of approximately $180 \kms$. On the other hand, if $\eta = 10^{-4}$ then the inferred velocity dispersion is approximately $560 \kms$, and it is marginal whether feedback from such a jet would influence the evolution of the host galaxy. For $\eta = 10^{-5}$, we expect no substantial feedback.

A number of patterns are immediately obvious from this diagram. Consider first the sequence of simulations CEFH. These have the same values of $\pism/k$ and warm cloud density, $\nwarmav$, and are ordered by decreasing jet power from $10^{46} \ergs$ down to $10^{43} \ergs$. All of these jets can disperse cloudy material for $\eta > \etacrit \approx 10^{-3.5}$ but for $\eta \lesssim 10^{-4}$ the jets would not strongly influence subsequent star formation. 

Next consider the BDG sequence. These simulations have a higher $\pism/k$ and the same ratio of cloud to ISM densities and, thus, the same filling factor. The clouds are denser and harder to move, and in this sequence the critical value of the jet power to Eddington ratio, $\eta$, is higher at around $10^{-3}$.

We have also conducted simulations, in which the filling factor is reduced by reducing the average density of the clouds for a fixed $\pism/k$. These are the sequences C$^\prime$E$^\prime$F$^\prime$ and B$^\prime$D$^\prime$G$^\prime$ plotted in Fig.~\ref{f:v-P}. At high jet powers, these lower filling factor simulations all exhibit higher values of $\etacrit$, that is, it is more difficult for the jet to disperse the clouds when the clouds are more porous. This is the result of the non-thermal plasma being able to escape more readily through a more porous medium. On the other hand, the simulations F$^\prime$ and G$^\prime$ at jet powers of $10^{44} \ergs$ occupy similar positions on the $\vrw$--$\Pjet$ diagram as their higher filling factor counterparts. In these cases the lower power jets do not break through as readily, and the confinement time of the non-thermal gas is sufficient to build up a large pressure, which accelerates the clouds. The results for even lower filling factors at all jet powers will be of interest but requires higher resolution simulations to prevent the clouds from becoming too pixelated.

Another aspect of jet feedback, which is readily interpreted using this diagram is the relative importance for feedback of jets of different power. Compare the points H and C in Fig.~\ref{f:v-P}. These points have approximately the same $\etacrit$. However, the \emph{maximum} value of the velocity dispersion, for which the jet in simulation H would satisfy the feedback criterion, $\vrw > \sigma$, is about $300 \kms$ whereas the corresponding value for C is about $2,300$.  Thus, we can make the qualitative remark that low-powered ($\Pjet \lesssim 10^{43} \ergs$) jets are relevant to feedback in less massive hosts, that is, galaxies in the early stages of hierarchical merging.

\subsection{Star formation in the over-pressured lobe}

Do stars form as a result of the excess pressure in the radio lobes? Let $p_0 = 10^{-9} p_{0,-9} \> \rm dyn \> cm^{-2}$ be the ambient pressure surrounding a sphere with central number density $\ncloud$ and temperature $10^4 \, T_4 \> \Kv$. If magnetic fields are neglected, considerations of star formation focus on the Bonnor-Ebert mass \citep{ebert55a,bonnor56a} $\MBE = 1.18 \> (kT/\bar{\mu} u)^2 G^{-3/2} p_0^{-1/2} \approx 2.0 \times 10^6 \> T_4^2 \, p_{0,-9}^{-1/2} \> \msolar$ or the
50\% larger critical mass $M_{\rm crit} \approx 3.0 \times 10^6 \> T_4^2 p_{0,-9}^{-1/2} \> \msolar$ derived from the virial theorem for a uniform density gravitating mass \citep{mccrea57a}. For a mass in excess of $M_{\rm crit}$, say, we expect that pressure driven gravitational instability will lead to star formation \citep{krumholz05a,antonuccio08a}. In the pre-jet phase of our highest pressure simulations ($p/k = 10^7 \> \Kv\cmq$) the critical mass is approximately $2.5 \times 10^6 \> \msolar$ compared to the mass, $10^6 \> \msolar$, of the largest clouds with radius, $\Rcloud \sim 25 \parsec$. Hence, at this stage, the mass of the largest clouds is less than the critical mass, although not by a great amount. What is the effect on star formation when the ambient pressure increases as a result of the formation of the high-pressured radio lobe? For example, in simulation F$^\prime$, the value of $p/k$, within the lobe rises from $10^7 \Kv\cmq$ to about $3 \times 10^{10} \Kv\cmq$, the critical mass decreases to about $5 \times 10^4 \> \msolar$ and jet induced starformation may be feasible. 

 However, the overpressured radio lobe is a highly turbulent environment and it is also helpful to compare the collapse timescale of the cloud, $\tcollapse \sim (4 \pi G \rhocloud)^{-1/2} \approx 2 \times 10^6 \, (\ncloud/300 \> \rm cm^{-3})^{-1/2} \yr$ to the ablation time scale $\tabl = 2 \Rcloud \vabl^{-1} \approx 8 \times 10^4 (\Rcloud/25\parsec) (\vabl/600\kms)^{-1}$, where $\vabl$ is the ablation velocity of cloud material. Clouds are destroyed on a time scales of a few $\times\tabl$. If we allow the cloud density to increase by a maximal factor of 100 as a result of radiative shocks and adopt fiducial values for the other parameters we have $\tcollapse \approx 2 \times 10^5 \yr$ and $\tabl \approx 8 \times 10^4 \yr$. For this extreme case one may expect star formation to occur. On the other hand, a factor of 100 increase in density is high so that we expect that most clouds would be dispersed before they could gravitationally collapse to form stars.  A similar conclusion was reached by \citet{antonuccio08a} in their study of the effect of radio-lobes driven by jets with powers ranging from $4 \times 10^{40}$ to $10^{46} \ergs$ interacting with a single cloud with a radius of $10\parsec$.
 
Given these estimates, under what circumstances would we expect \emph{jet-induced} rather that \emph{jet-inhibited} star formation? First, if the temperature of the clouds were lower than $10^4 \> \rm K$ the critical mass for gravitational instability ($\propto T^2$) would also be lower. Second, we can also speculate that the clouds towards the edge of the radio lobe would not be so strongly affected by the lower level of turbulence in that region but they are still initially affected by the overpressure of the radio lobe bow-shock. Therefore it is possible that there may be some star formation in these regions. These issues are beyond the scope of this paper but they should be of interest in future work.

\subsection{Efficiency of kinetic energy transfer}

\begin{figure*}
  \epsscale{1.00}
  \figurenum{6}
  \plotone{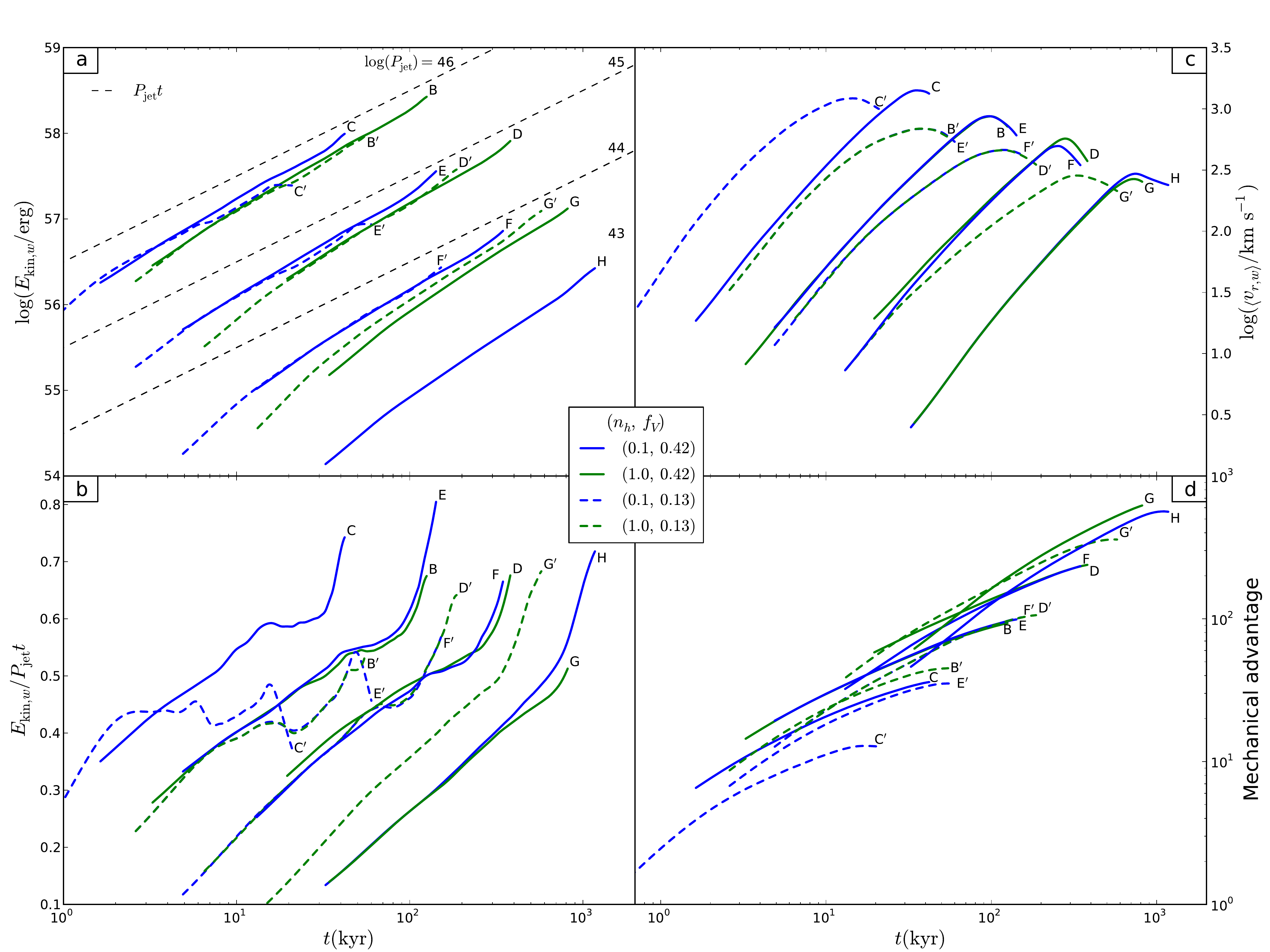}
  \caption{Evolution of various quantities in simulations B--H as functions of time. In all panels, the blue and green curves refer to simulations, for which $\nhot = 1.0$ and $\nhot=0.1$, respectively, the solid curves refer to a volume filling factor $\fvol = 0.42$, and the dashed curves to $\fvol = 0.13$. (a) Kinetic energy of clouds. The dashed lines indicate the energy delivered by the jets with the jet power indicated on each curve. (b) Efficiency of energy transfer defined as the ratio of kinetic energy of clouds to the total energy delivered by the jet. (c) Density-averaged radial velocity of clouds. (d) Mechanical advantage, which is the ratio of the total momentum delivered by the jet to the total radial momentum in clouds.}
  \label{f:combi}
\end{figure*}

Panels a and b in Fig.~\ref{f:combi} show the kinetic energy imparted to the clouds in each simulation and the efficiency of the transfer of energy from the jet to the clouds, respectively. In each simulation the efficiency is quite high and, except for runs C$^\prime$ and E$^\prime$, the efficiency is still increasing by the end of the simulation. All curves show a slight dip or inflection before steeply rising; this feature occurs at the point where the jet plasma breaks out of the confining clouds. (In the simulations of lower powered jets, the break out is smoothed out because the non-thermal plasma breaks out everywhere almost simultaneously.)

These curves show a decrease in efficiency with decreasing value of the ratio of jet power to cloud density ($\Pjet/\nwarmav$). From the curves of $\vrw$ as a function of time shown in panel c of Fig.~\ref{f:combi}, we see that the maxima in $\vrw$ occur just before jet break-out, which releases some of the pressure in the quasi-spherical bubble. Compare the similarity of the evolution of $\Ekinw/\Pjet$ in B and E, D and F, G and H, B$^\prime$ and E$^\prime$ and D$^\prime$ and F$^\prime$. These pairs of runs have the same values of $\Pjet/\nwarmav$ and exhibit comparable maximum radial velocity dispersion.

\subsection{Momentum transfer}

The rate of momentum transfer from the jet to the clouds can be an issue in this field, to which \citet{krause10a} drew attention. For example, the range of cloud and jet parameters presented in \citet{holt06a} indicates that the cloud momentum flux may exceed that of the jet. Hence, it is useful to examine the momentum budget to see what can be expected from the observations.  Momentum is not as straightforward as energy since the momentum in a volume surrounding a moving cloud is affected by the pressure (and magnetic stresses) integrated over the bounding surface of that volume. In this case the shocked jet gas provides a large pressure as a result of the \emph{energy} flux in the jet.  Therefore, in panel d of Fig.~\ref{f:combi}, we present the mechanical advantage (the ratio of the momentum in clouds to the jet momentum) as a function of time in each simulation. In each case the mechanical advantage exceeds unity. Note that the greatest mechanical advantage is achieved by the lowest power jets. Low powered jets "punch through" the warm clouds more slowly, and the shocked jet gas spreads more effectively through the volume of warm clouds and has a greater relative effect on the cloud momentum. 

\subsection{Late-time evolution}

In each of the simulations, the jet plasma, after a sufficiently long time, escapes the cloud region. The more powerful the jet is, and the lighter the clouds are, the more centrally collimated the outbreak of the plasma is. We end the simulations when the jet plasma reaches the domain boundaries, and boundary effects become noticeable. We may, nonetheless, speculate on the subsequent evolution of the system.

In the highest power jet simulations, we see the jet breaking through along its principal propagation axis ($x=0$) and piercing the energy bubble that was inflated during the flood-and-channel phase of evolution. Except for runs G and H, the jets in all our simulations will eventually emerge in a similar fashion, inflating a second, more classically shaped radio lobe beyond the first quasi-spherical bubble to a distance depending on the jet power. In runs G and H, the jet may be trapped indefinitely in the cloud region. While the common notion is that hierarchical radio lobe structures are produced by recurrent jet activity \citep{SJ2009}, we note that they may also be produced by the randomized emergence of radio plasma from an inhomogeneous, porous ISM.

After the jet escapes along its principal propagation axis, the efficiencies in transfer of momentum and energy to the clouds drop quickly. This stage is already seen in runs C$^\prime$ and E$^\prime$, and will occur at later times in the other runs (except for G and H, if the jets in those runs remains trapped for the duration of their activity). While the clouds in our simulations are dispersed to high velocities, the host galaxy will retain most of the cloud material, except for a diffuse component comprising the high velocity tail in the distribution, which is concentrated along the jet axis near the region where the central jet breaks out. By the end of the runs, the clouds in all simulations have reached and passed their maximum value of $\vrw$.

\section{Discussion}
\label{s:discussion}

The suite of simulations presented in this paper strongly reinforce the importance of inhomogeneity in the consideration of jet interactions with the ISM \citep{saxton05a,sutherland07a}. Inhomogeneity has several important consequences: (1) It affects the early morphology of the radio source as a result of the interaction of the jet and lobe with the obstructing clouds. (2) The radio source affects a much larger volume of the host galaxy because of the channeling of the jet flow in different directions.  (3) The shocked jet clouds are left in the wake of the non-thermal plasma and, as first noted by \citet{sutherland07a}, would continue to emit shock excited line emission with the shocks driven by the high pressure gas in the non-thermal cocoon; this shock-excited emission is in addition to the emission that may be driven by photoionization by the nucleus. (4) The porosity of dense gas determines the ease with which a jet in a given host can disperse this gas, which is a potential source of new stars. In powerful sources, higher porosity gas is less easily dispersed but this trend is not evident for lower-powered ($\sim 10^{44} \ergs$) jets. (5) Jets of all powers can exert a considerable feedback effect on their host galaxies, although lower-powered jets only play a role in the lower velocity dispersion hosts. Brighter galaxies require more powerful jets to disperse dense clouds. (6) The efficiency of transfer of kinetic energy from the jet to the dense gas is high. (7) The efficiency of transfer of momentum to the clouds is also high with mechanical advantages considerably exceeding unity.

Inhomogeneity is therefore crucial when considering AGN feedback on the kiloparsec scale, both for the interpretation of radio and optical emission-line morphology in radio sources, which may be generating feedback, and for incorporating the effect of jet-mediated feedback on host galaxies of different size into large scale simulations, in which the resolution $\gtrsim 1 \kpc$.

An important conclusion from these simulations is that jets with Eddington efficiency $\eta \lesssim 10^{-4}$ are unlikely to have an effect on evolving galaxies when the pre-starforming gas exists in the form of clouds, which are relatively dense and cool compared to the hot ISM. This critical value of $\eta$ is relevant for clouds with a high filling factor of 0.42 and a value of $p/k = 10^6$ for the ISM. We have shown that when the filling factor decreases or the pressure of the ISM increases the critical value of $\eta$ increases, and values of $\etacrit \sim 10^{-3} - 10^{-2}$ are not unrealistic. The precise values of $\etacrit$ will have to await further higher resolution simulations with high porosity dense gas.



\acknowledgments
This research was undertaken on the NCI National Facility at the Australian National University. The software used in this work was in part developed by the DOE-supported ASC / Alliance Center for Astrophysical Thermonuclear Flashes at the University of Chicago. We are grateful to Prof. Mitchell Begelman for detailed comments on an initial version of the manuscript and to Dr. Ralph Sutherland who provided revised cooling functions, incorporating the latest solar abundances, for this work. We are grateful to our anonymous referee for a thorough reading of the manuscript and for a number of constructive comments, which have improved the presentation.








\begin{thebibliography}{}

\bibitem[\protect\citeauthoryear{{Antonuccio-Delogu} \&
  {Silk}}{{Antonuccio-Delogu} \& {Silk}}{2008}]{antonuccio08a}
{Antonuccio-Delogu}, V. \& {Silk}, J. 2008, \mnras,{ 389}, 1750--1762
\bibitem[\protect\citeauthoryear{{Asplund}, {Grevesse}, \& {Sauval}}{{Asplund}
  et~al.}{2005}]{AGS2005}
{Asplund}, M., {Grevesse}, N., \& {Sauval}, A.~J. 2005, September), Volume 336
  of Astronomical Society of the Pacific Conference Series 25--+
\bibitem[\protect\citeauthoryear{{Bender} \& {Saglia}}{{Bender} \&
  {Saglia}}{1999}]{bender99a}
{Bender}, R. \& {Saglia}, R.~P. 1999, August), in Galaxy Dynamics - A Rutgers
  Symposium, ed. {D.~R.~Merritt, M.~Valluri, \& J.~A.~Sellwood}, Volume 182 of
  Astronomical Society of the Pacific Conference Series 113--+
\bibitem[\protect\citeauthoryear{{Bonnor}}{{Bonnor}}{1956}]{bonnor56a}
{Bonnor}, W.~B. 1956, \mnras,{ 116}, 351--359
\bibitem[\protect\citeauthoryear{{Booth} \& {Schaye}}{{Booth} \&
  {Schaye}}{2009a}]{booth09b}
{Booth}, C.~M. \& {Schaye}, J. 2009a, \mnras,{ 398}, 53--74
\bibitem[\protect\citeauthoryear{{Booth} \& {Schaye}}{{Booth} \&
  {Schaye}}{2009b}]{booth09a}
{Booth}, C.~M. \& {Schaye}, J.ArXiv e-prints,
\bibitem[\protect\citeauthoryear{{Cole} et~al.}{{Cole} et~al.}{2001}]{cole01a}
{Cole}, S., {et~al.} 2001, \mnras,{ 326}, 255--273
\bibitem[\protect\citeauthoryear{{Croton} et~al.}{{Croton}
  et~al.}{2006}]{croton06a}
{Croton}, D.~J., {et~al.} 2006, MNRAS,{ 365}, 11--28
\bibitem[\protect\citeauthoryear{{Dopita} et~al.}{{Dopita}
  et~al.}{2004}]{dopita04a}
{Dopita}, M.~A., {Fischera}, J., {Groves}, B., {Sutherland}, R.~S., {Kewley},
  L.~J., {Tuffs}, R., {Popescu}, C., \& {Leitherer}, C. 2004, in The Interplay
  Among Black Holes, Stars and ISM in Galactic Nuclei, ed.
  {T.~Storchi-Bergmann, L.~C.~Ho, \& H.~R.~Schmitt}, Volume 222 of IAU
  Symposium 313--314
\bibitem[\protect\citeauthoryear{{Ebert}}{{Ebert}}{1955}]{ebert55a}
{Ebert}, R. 1955, Zeitschrift fur Astrophysik,{ 37}, 217--232
\bibitem[\protect\citeauthoryear{{Fabian}}{{Fabian}}{1999}]{fabian99a}
{Fabian}, A.~C. 1999, \mnras,{ 308}, L39--L43
\bibitem[\protect\citeauthoryear{{Fabian} et~al.}{{Fabian}
  et~al.}{2003}]{fabian03a}
{Fabian}, A.~C., {Sanders}, J.~S., {Allen}, S.~W., {Crawford}, C.~S.,
  {Iwasawa}, K., {Johnstone}, R.~M., {Schmidt}, R.~W., \& {Taylor}, G.~B. 2003,
  September), \mnras,{ 344}, L43--L47
\bibitem[\protect\citeauthoryear{{Federrath} et~al.}{{Federrath}
  et~al.}{2010}]{Federrath-etal2010}
{Federrath}, C., {Roman-Duval}, J., {Klessen}, R.~S., {Schmidt}, W., \& {Mac
  Low}, {M.-M.} 2010, March), \aap,{ 512}, A81+
\bibitem[\protect\citeauthoryear{{Fischera}, {Dopita}, \&
  {Sutherland}}{{Fischera} et~al.}{2003}]{fischera03a}
{Fischera}, J., {Dopita}, M.~A., \& {Sutherland}, R.~S. 2003, ApJL,{ 599},
  L21--L24
\bibitem[\protect\citeauthoryear{{Fryxell} et~al.}{{Fryxell}
  et~al.}{2000}]{Fryxell-etal2000}
{Fryxell}, B., {et~al.} 2000, November), \apjs,{ 131}, 273--334
\bibitem[\protect\citeauthoryear{{Gebhardt} et~al.}{{Gebhardt}
  et~al.}{2000}]{gebhardt00a}
{Gebhardt}, K., {et~al.} 2000, ApJL,{ 539}, L13
\bibitem[\protect\citeauthoryear{{Holt} et~al.}{{Holt} et~al.}{2006}]{holt06a}
{Holt}, J., {Tadhunter}, C., {Morganti}, R., {Bellamy}, M., {Gonz{\'a}lez
  Delgado}, R.~M., {Tzioumis}, A., \& {Inskip}, K.~J. 2006, \mnras,{ 370},
  1633--1650
\bibitem[\protect\citeauthoryear{{Huang} et~al.}{{Huang}
  et~al.}{2003}]{huang03a}
{Huang}, {J.-S.}, {Glazebrook}, K., {Cowie}, L.~L., \& {Tinney}, C. 2003,
  \apj,{ 584}, 203--209
\bibitem[\protect\citeauthoryear{{King}}{{King}}{2005}]{king05a}
{King}, A. 2005, \apjl,{ 635}, L121--L123
\bibitem[\protect\citeauthoryear{{Krause} \& {Gaibler}}{{Krause} \&
  {Gaibler}}{2010}]{krause10a}
{Krause}, M. \& {Gaibler}, V. 2010, in AGN Feedback in Galaxy Formation, ed.
  V.~Antonuccio-Delogu \& J.~Silk in press
\bibitem[\protect\citeauthoryear{{Krumholz} \& {McKee}}{{Krumholz} \&
  {McKee}}{2005}]{krumholz05a}
{Krumholz}, M.~R. \& {McKee}, C.~F. 2005, \apj,{ 630}, 250--268
\bibitem[\protect\citeauthoryear{{Lewis} \& {Austin}}{{Lewis} \&
  {Austin}}{2002}]{LA2002}
{Lewis}, G.~M. \& {Austin}, P.~H. 2002, in 11th Conference on Atmospheric
  Radiation, ed. G.~H. {Smith} \& J.~P. {Brodie}, American Meteorological
  Society Conference Series 123--126
\bibitem[\protect\citeauthoryear{{Madau} et~al.}{{Madau}
  et~al.}{1996}]{madau96a}
{Madau}, P., {Ferguson}, H.~C., {Dickinson}, M.~E., {Giavalisco}, M.,
  {Steidel}, C.~C., \& {Fruchter}, A. 1996, \mnras,{ 283}, 1388--1404
\bibitem[\protect\citeauthoryear{{Magorrian} et~al.}{{Magorrian}
  et~al.}{1998}]{magorrian98a}
{Magorrian}, J., {et~al.} 1998, AJ,{ 115}, 2285--2305
\bibitem[\protect\citeauthoryear{{Mathews} \& {Brighenti}}{{Mathews} \&
  {Brighenti}}{2003}]{MB2003}
{Mathews}, W.~G. \& {Brighenti}, F. 2003, \araa,{ 41}, 191--239
\bibitem[\protect\citeauthoryear{{McCrea}}{{McCrea}}{1957}]{mccrea57a}
{McCrea}, W.~H. 1957, \mnras,{ 117}, 562
\bibitem[\protect\citeauthoryear{{McNamara} et~al.}{{McNamara}
  et~al.}{2005}]{mcnamara05a}
{McNamara}, B.~R., {Nulsen}, P.~E.~J., {Wise}, M.~W., {Rafferty}, D.~A.,
  {Carilli}, C., {Sarazin}, C.~L., \& {Blanton}, E.~L. 2005, January), \nat,{
  433}, 45--47
\bibitem[\protect\citeauthoryear{{Mignone} \& {Bodo}}{{Mignone} \&
  {Bodo}}{2005}]{MB2005}
{Mignone}, A. \& {Bodo}, G. 2005, November), \mnras,{ 364}, 126--136
\bibitem[\protect\citeauthoryear{{Mignone}, {Plewa}, \& {Bodo}}{{Mignone}
  et~al.}{2005}]{MPB2005}
{Mignone}, A., {Plewa}, T., \& {Bodo}, G. 2005, September), \apjs,{ 160},
  199--219
\bibitem[\protect\citeauthoryear{{Mizuno}, {Hardee}, \& {Nishikawa}}{{Mizuno}
  et~al.}{2007}]{MHN2007}
{Mizuno}, Y., {Hardee}, P., \& {Nishikawa}, {K.-I.} 2007, June), \apj,{ 662},
  835--850
\bibitem[\protect\citeauthoryear{{Morganti} et~al.}{{Morganti}
  et~al.}{2010}]{morganti10a}
{Morganti}, R., {Holt}, J., {Tadhunter}, C., \& {Oosterloo}, T.ArXiv e-prints,
\bibitem[\protect\citeauthoryear{{Nesvadba} et~al.}{{Nesvadba}
  et~al.}{2009}]{nesvadba09b}
{Nesvadba}, N.~P.~H., {et~al.} 2009, \mnras,{ 395}, L16--L20
\bibitem[\protect\citeauthoryear{{Norberg} et~al.}{{Norberg}
  et~al.}{2002}]{norberg02a}
{Norberg}, P., {et~al.} 2002, \mnras,{ 336}, 907--931
\bibitem[\protect\citeauthoryear{{Nulsen} et~al.}{{Nulsen}
  et~al.}{2009}]{nulsen09a}
{Nulsen}, P., {Jones}, C., {Forman}, W., {Churazov}, E., {McNamara}, B.,
  {David}, L., \& {Murray}, S.ArXiv e-prints,
\bibitem[\protect\citeauthoryear{{Saikia} \& {Jamrozy}}{{Saikia} \&
  {Jamrozy}}{2009}]{SJ2009}
{Saikia}, D.~J. \& {Jamrozy}, M. 2009, September), Bulletin of the Astronomical
  Society of India,{ 37}, 63--89
\bibitem[\protect\citeauthoryear{{Saxton} et~al.}{{Saxton}
  et~al.}{2005}]{saxton05a}
{Saxton}, C.~J., {Bicknell}, G.~V., {Sutherland}, R.~S., \& {Midgley}, S. 2005,
  MNRAS,{ 359}, 781--800
\bibitem[\protect\citeauthoryear{{Schaye} et~al.}{{Schaye}
  et~al.}{2010}]{schaye10a}
{Schaye}, J., {et~al.} 2010, \mnras,{ 402}, 1536--1560
\bibitem[\protect\citeauthoryear{{Shaver} et~al.}{{Shaver}
  et~al.}{1996}]{shaver96a}
{Shaver}, P.~A., {Wall}, J.~V., {Kellermann}, K.~I., {Jackson}, C.~A., \&
  {Hawkins}, M.~R.~S. 1996, \nat,{ 384}, 439--441
\bibitem[\protect\citeauthoryear{{Silk}}{{Silk}}{1997}]{Silk1997}
{Silk}, J. 1997, May), \apj,{ 481}, 703--+
\bibitem[\protect\citeauthoryear{{Silk}}{{Silk}}{2001}]{Silk2001}
{Silk}, J. 2001, June), \mnras,{ 324}, 313--318
\bibitem[\protect\citeauthoryear{{Silk} \& {Norman}}{{Silk} \&
  {Norman}}{2009}]{SN2009}
{Silk}, J. \& {Norman}, C. 2009, July), \apj,{ 700}, 262--275
\bibitem[\protect\citeauthoryear{{Silk} \& {Rees}}{{Silk} \&
  {Rees}}{1998}]{silk98a}
{Silk}, J. \& {Rees}, M.~J. 1998, A\&A,{ 331}, L1--L4
\bibitem[\protect\citeauthoryear{{Springel}}{{Springel}}{2005}]{springel05a}
{Springel}, V. 2005, MNRAS,{ 364}, 1105--1134
\bibitem[\protect\citeauthoryear{{Springel} \& {Hernquist}}{{Springel} \&
  {Hernquist}}{2003a}]{springel03b}
{Springel}, V. \& {Hernquist}, L. 2003a, \mnras,{ 339}, 289--311
\bibitem[\protect\citeauthoryear{{Springel} \& {Hernquist}}{{Springel} \&
  {Hernquist}}{2003b}]{springel03a}
{Springel}, V. \& {Hernquist}, L. 2003b, \mnras,{ 339}, 312--334
\bibitem[\protect\citeauthoryear{{Sutherland} \& {Bicknell}}{{Sutherland} \&
  {Bicknell}}{2007}]{sutherland07a}
{Sutherland}, R.~S. \& {Bicknell}, G.~V. 2007, ApJS,{ 173}, 37--69
\bibitem[\protect\citeauthoryear{{Sutherland}, {Bicknell}, \&
  {Dopita}}{{Sutherland} et~al.}{2003}]{sutherland03b}
{Sutherland}, R.~S., {Bicknell}, G.~V., \& {Dopita}, M.~A. 2003, ApJ,{ 591},
  238--257
\bibitem[\protect\citeauthoryear{Sutherland \& Dopita}{Sutherland \&
  Dopita}{1993}]{sutherland93c}
Sutherland, R.~S. \& Dopita, M.~A. 1993, ApJS,{ 88}, 253
\bibitem[\protect\citeauthoryear{{Tremaine} et~al.}{{Tremaine}
  et~al.}{2002}]{tremaine02a}
{Tremaine}, S., {et~al.} 2002, ApJ,{ 574}, 740--753
\end{thebibliography}
\bibliographystyle{apjv2}



\end{document}